\documentclass[aps,prl,twocolumn,superscriptaddress,showpacs]{revtex4}
\usepackage{graphicx}
\draft

\begin{document}

\title{Numerical simulations of a ballistic spin interferometer
with Rashba spin-orbital interaction}

\author{Zhenyue Zhu}
\affiliation{Department of Physics, Oklahoma State University,
Stillwater, Oklahoma 74078 }

\author{Qing-feng Sun}
\affiliation{Beijing National Lab for Condensed Matter Physics and
Institute of Physics, The Chinese Academy of Sciences, Beijing
100080, P.R. China}

\author{Bin Chen}
\affiliation{Department of Physics, Oklahoma State University,
Stillwater, Oklahoma 74078 } \affiliation{Department of
Physics,Hangzhou Teachers College, Hangzhou 310036, P.R. China}

\author{X. C. Xie}
\affiliation{Department of Physics, Oklahoma State University,
Stillwater, Oklahoma 74078 } \affiliation{Beijing National Lab for
Condensed Matter Physics and Institute of Physics, The Chinese
Academy of Sciences, Beijing 100080, P.R. China}

\date{\today}

\begin{abstract}
We numerically investigate the transport behavior of a quasi
one-dimension (1D) square loop device containing the Rashba
spin-orbital interaction in the presence of a magnetic flux. The
conductance versus the magnetic field shows the
Al'tshuler-Aronov-Spivak (AAS) and Aharonov-Bohm (AB)
oscillations. We focus on the oscillatory amplitudes, and find
that both of them are strongly dependent on the spin precession
angle (i.e. the strength of the spin-orbit interaction) and
exhibit no-periodic oscillations, which are well in agreement with
a recent experiment by Koga \textit{et al.} \cite{ref8}. However,
our numerical results for the ideal 1D square loop device for the
node positions of the amplitudes of the AB and AAS oscillations
are found to be of some discrepancies comparing with quasi-1D
square loop with a finite width. In the presence of disorder and
taking the disorder ensemble average, the AB oscillation in the
conductance will disappear, while the time-reversal symmetric AAS
oscillation still remains. Furthermore, the node positions of the
AAS oscillatory amplitude remains the same.

\end{abstract}


\maketitle

The control over the transport properties of electron spins has
gained much attention recently\cite{ref1,ref2,ref3}. A new
sub-discipline of condensed matter physics, spintronics, is
emerging rapidly and generating great interests. The promising
application of spintronics widely lies in, e.g. information
technology\cite{ref4}, the spin electron apparatus\cite{ref5}, and
so on. Utilizing the spin orbital (SO) interaction to manipulate
the spin degrees of freedoms of electrons has been advanced.
Several spin devices have also been theoretically designed.
Experimentally, the strength of Rashba SO interaction has
successfully been controlled by the applied gate voltage or with
some specific design of heterostructures \cite{ref6,ref7}.

Very recently, Koga \textit{et al.} used a nanolithographically
defined square loop array in the two dimensional electron gases to
experimentally demonstrate a gate-controlled electron spin
interference \cite{ref8,addref1}. They observed that the amplitude
of the Al'tshuler-Aronov-Spivak (AAS) type oscillation \cite{ref9}
of the conductance versus the magnetic field $B$ depends strongly
on the gate voltage, or the sheet carrier density. This implies
that electron spin precession angle $\theta$, caused by the Rashba
SO interaction \cite{ref10}, is gate-controllable, and can be
tuned more than $0.75\pi$. In fact, to experimentally realize a
large controllability of the spin precession angle $\theta$ is a
very important issue in spintronics, because it is indispensable
for realizing some spin devices, e.g. the spin field effect
transistor \cite{ref5}.

Before the experimental work \cite{ref8}, some theoretical works
have investigated the transport behavior of a ring with the SO
interaction \cite{addref2,addref3,addref4}. In particular, a
theoretical prediction of the amplitude of the AAS oscillation has
been made by the same group \cite{ref11}. Although both the
experimental and theoretical results are similar, there are some
discrepancies. The theoretical system is an ideal perfect one
dimension (1D) square loop with only one terminal contacted outer,
for which the backscattering probability was calculated. In the
experiment, the device is a quasi-1D square loop (i.e. each arm of
the loop has a finite width) with multi terminals, and the
transport conductance instead of the backscattering probability
was measured. The authors argued that the results for both systems
should be similar, however, no through investigation was given.
Moreover, in the experimental device, the first order oscillation
is the Aharonov-Bohm (AB) oscillation with the period $\phi_0$
($\phi_0=h/e$), which is never studied.

In this paper, we numerically simulate the experimental set-up by
investigating the amplitude of the AB and the AAS oscillation. We
consider a quasi-1D square loop having the Rashba SO interaction
(see Fig.1a), which represents one of the cells in the
experimental setup by Koga \textit{et al.} \cite{ref8}. The
conductance is first calculated from the Landauer-B\"{u}ttiker
formula. Afterwards, by applying the same data analysis process as
in the experiment, i.e. the fast Fourier transform (FFT) and the
inverse fast Fourier transform (IFFT), the amplitude of the AAS
and AB oscillations of conductance are obtained. The numerically
calculated node positions of amplitude of AAS oscillation in
quasi-1D device are of some departures ($\sim 0.1\pi$) from the
theoretical results for the ideal 1D device. The difference is
larger for a loop with a wider arm. Moreover, the node positions
of the amplitude of the AB oscillation are also studied. Finally,
we discuss how the node positions affected by the Dresselhaus SO
interaction and the disorder, and find that the Dresselhaus SO
interaction can strongly shift the node positions, but the
disorder has almost no effect on the node positions.

\begin{figure}[ht]
\begin{center}
\includegraphics[height=8.5cm,angle=270]{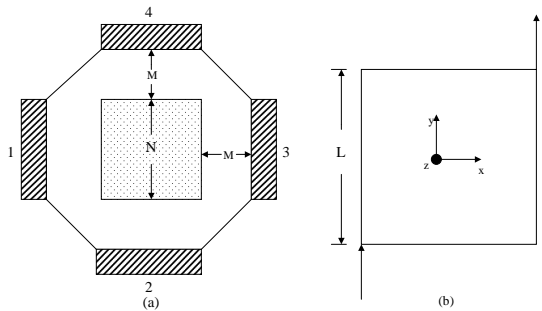}
\end{center}
\caption{ Sketchy illustration for a quasi-1D (a) and an ideal 1D
(b) square loop devices. In (a), the potential in the center
region is assumed to be very high so that there are no electrons
there. }
\end{figure}

Before investigating the quasi-1D loop system, we first study an
ideal 1D square loop model attached with two 1D leads on its two
opposite corners (shown in Fig.1b). The goal is to find
analytically the node positions of the amplitude of AB or AAS
oscillation in the two-terminal system. The Rashba SO interaction
exists only in the loop, but absent in the two leads.\cite{note11}
An incident wave $\psi_i$ is splitted at the lower left corner.
And the counterclockwise (CCW) and clockwise (CW) wave
ballistically travelling along two sides of square loop merge at
the right upper corner. $\psi_{o}$ describes the output wave
function of the electron at the right upper corner. If only the
first order tunneling process is considered, $\psi_{o}$ can be
obtained as:
\begin{eqnarray}
\psi_{o}= \frac{1}{2} \left( e^{-i\phi/2}R_{-x}(\theta)
R_{y}(\theta) +  e^{i\phi/2}R_{y}(\theta) R_{-x}(\theta) \right)
\psi_i.
\end{eqnarray}
where $\phi =2\pi \Phi/\phi_0$ with the magnetic flux $\Phi$. The
rotation operator $R_{\hat{r}}(\theta)$ is defined as:\cite{ref11}
\begin{equation}
R_{\hat{r}}(\theta)=\textbf{I}\cos(\theta/2)-i\hat{r}\cdot{\bf
\sigma}\sin(\theta/2),
\end{equation}
and $\theta= 2 m^*\alpha_R L /\hbar^2 $ is the spin precession
angle. Here $L$ is the side length of the 1D square loop and
$\alpha_R$ is the strength of the Rashba SO interaction. To
consider that the incident electron is spin unpolarized, the
output probability of the electron is
$\overline{<\psi_o|\psi_o>}$, and it's given by
\begin{equation}
\overline{<\psi_o|\psi_o>}=\frac{1}{2}+ A(\theta) \cos\phi ,
\end{equation}
where $A(\theta)=1/4(\sin^2\theta+2\cos\theta)$ is the amplitude
of the AB oscillation. Then the node positions (marked by
$\theta^*$) can be obtained easily, and they are $0.636\pi$ and
$1.364\pi$, etc. The amplitudes of the higher order oscillation
(including the AAS oscillation) are zero here, because the higher
order tunnelling process has been neglected in Eq.(1).

Next, we study the model of a quasi-1D square loop sketched in
Fig.1a. With the model, no electron exists in the center region of
the loop, which can be experimentally realized by the etching
technology or the deposited metal split gate. Four leads
symbolized as hatched regions in Fig.1a are attached to the four
sides of the system. Rashba SO interaction and magnetic field only
exist in the quasi loop, and four leads are ideal without SO
interaction \cite{note11}. The width of the lead is N, the channel
width is M, and the side length for the quasi-1D square loop is
L=M+N. This quasi-1D model is identical with each cell of the
experimental setup in the Ref. \cite{ref8}. In their experiment,
they applied a square loop array to determine the sheet
conductance which is necessary to diminish the AB oscillation and
the universal conductance fluctuations.

The Hamiltonian for the quasi-1D model is described as a discrete
lattice tight binding model in our numerical calculations. We
choose a symmetric gauge and the vector potential $\vec A(x,y) =
B(-y/2, x/2, 0)$, with the lattice spacing as the length unit. Let
$a_{j\sigma}$ ($a^{\dag}_{j\sigma}$) be the annihilation
(creation) operator of an electron with its spin $\sigma$ at the
lattice site $j$, and $a^{\dag}_j = (a^{\dag}_{j \uparrow},
a^{\dag}_{j\downarrow})$. Then the tight binding model Hamiltonian
for the system can be written as:

\begin{eqnarray}
H &=&
-\sum_{j\tau\sigma}(t_{j\tau}a^{\dag}_{j,\sigma}a_{j+\tau,\sigma}+h.c)
 + \sum_{j\sigma}w_{j}a^{\dag}_{j\sigma}a_{j\sigma}\nonumber\\
&+&  \sum_{j}[i\hbar(\lambda_{jx}a^{\dag}_{j} \sigma_y a_{j+\hat
x} -\lambda_{jy} a^{\dag}_{j} \sigma_x a_{j+\hat y}) + h.c.]
\end{eqnarray}
In the above equation, $\sigma_{x(y)}$ are Pauli matrices, and
\begin{eqnarray}
t_{j\tau} &=& t\cdot e^{i\frac{e}{\hbar}\vec{A}\cdot\tau
},\\
\lambda_{j\tau}&=& \lambda \cdot
e^{i\frac{e}{\hbar}\vec{A}\cdot\tau},
\end{eqnarray}
where $t=\frac{\hbar^2}{2m^* a^2}$ is the nearest neighbor hopping
element with the lattice space $a$, $\lambda=\frac{\alpha_R}{2a}$
describes the strength of the Rashba SO interaction, and $\tau
=\pm \hat x, \pm \hat y$. We did not consider the Zeeman splitting
energy, because the magnetic field in the experiment is so weak
that the Zeeman term is negligibly small compared with other
energies, e.g. $t$, the Rashba splitting energy, etc. The disorder
potential at each site is not considered right now. Thus we set
$w_j=0$ in the Hamiltonian as energy zero point. The tight binding
Hamiltonian for the ideal 1D square loop system is also given with
the same approximations.

The current $I_p$ from the lead $p$ ($p=1$, 2, 3, and $4$) flowing
into the square loop can be calculated by the
Landauer-B\"{u}ttiker formulae\cite{ref12,ref13}:
\begin{equation}
I_p= \frac{2e^2}{h}\sum_q T_{pq}(E)[V_p-V_q].
\end{equation}
where $V_q$ is the voltage applied on the lead $q$, $T_{pq}(E)$ is
the transmission probability from the lead $q$ to the lead $p$,
and $E$ is the electron Fermi energy. The temperature is set to be
zero, because the thermal energy $k_BT$ is much smaller than the
other energy scales in the experiment. The transmission
probability $T_{pq}$ is determined by
$T_{pq}=Tr[\Gamma_pG^R\Gamma_qG^A]$ \cite{ref14,ref15}. Here
$\Gamma_p=i[\Sigma^R_p-\Sigma^A_p]$, and $G^{R(A)}(E)$ is the
retarded (advanced) Green function given by:
\begin{equation}
G^{R}(E)= (G^{A}(E))^\dagger =\frac{1}{E -H_c-\sum\limits_{p=1}^4
\Sigma^R_p},
\end{equation}
where $\Sigma_p^{R(A)}$ is the retarded (advanced) self-energy of
lead $p$ and $H_c$ is the single particle Hamiltonian given by Eq.
(4) for the isolated finite-size system at the center. In the
following numerical calculations, the lead's voltages $V_p$ are
set as: $V_1= V/2$ and $V_3=-V/2$, i.e. a longitudinal bias is
added in the $x$ direction. The transverse lead-2 and lead-4 act
as the voltage probes, and their voltages $V_2$ and $V_4$ are
calculated from the condition $I_2=I_4=0$. Finally, the
longitudinal conductance $\sigma$ is obtained as: $\sigma =I_1/V =
-I_3/V$. In the numerical calculations, we let $t=1$ to be the
energy unit.
\begin{figure}[ht]
\begin{center}
\includegraphics[height=8.5cm,angle=270]{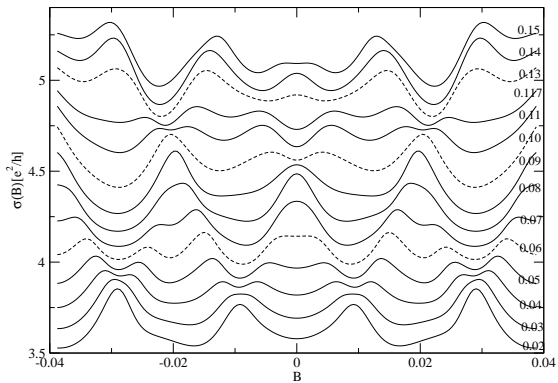}
\end{center}
\caption{The longitudinal conductance $\sigma(B)$ versus the
magnetic field $B$ for the quasi-1D device at different Rashba
interaction $\lambda$, with the parameters $N=12$, $M=6$, and the
Fermi energy $E=-2.0$. The value of $\lambda$ is indicated with
each curve. B is in the unit of $\frac{a^2e}{\hbar}$. The plotted
curves are shifted along the y-axis for comparison.  }
\end{figure}

The longitudinal conductance $\sigma$ of the quasi-1D square loop
system with N=12 and M=6 versus magnetic field $B$ at different
spin orbital interaction $\lambda$ are shown in Fig.2. These
conductance curves clearly exhibit the existence of the AB
oscillation with the period $\phi_0$ and the AAS oscillation with
the period $\phi_0/2$. Since each arm of the loop has a certain
width, these oscillations are not exactly periodic. Their
oscillatory amplitudes at $B=0$ can be figured out by using the
FFT. From the graph, we also find the alternative changing of peak
and dip feature in $\sigma(B)$ at B=0 with increasing of the
Rashba interaction. This phenomenon is consistent with the
experimental results \cite{ref8}.

Then for each curve of the conductance in Fig.2 with a fixed value
of the Rashba strength $\lambda$, we can employ the FFT and IFFT
techniques to extract out the amplitudes of the AB and AAS
oscillations at zero magnetic field. The data analysis is
identical with the experimental procedure. Since
$\lambda=\frac{\alpha_R}{2a}$ and $t= \frac{\hbar^2}{2m^* a^2}$,
the Rashba strength $\lambda$ and the spin precession angle
$\theta$ is related by $\theta=\frac{2L}{a t}\lambda$. Fig.3 shows
the amplitudes $\sigma(B=0)$ of the AB and the AAS oscillations at
$B=0$ versus the spin precession angle $\theta$ for different
sizes of the device. The node positions $\theta^*$ for both AB and
AAS oscillations are indicated in each graph. For comparison, the
amplitudes of the AB and the AAS oscillations of the exact 1D
square loop system (see Fig.1b) are also calculated by using the
tight binding model and the Landauer-B\"{u}ttiker formulae, in
which the higher order processions and the reflecting processions
have all been included (see Fig.4).
\begin{figure}[ht]
\begin{center}
\includegraphics[height=8.5cm,angle=270]{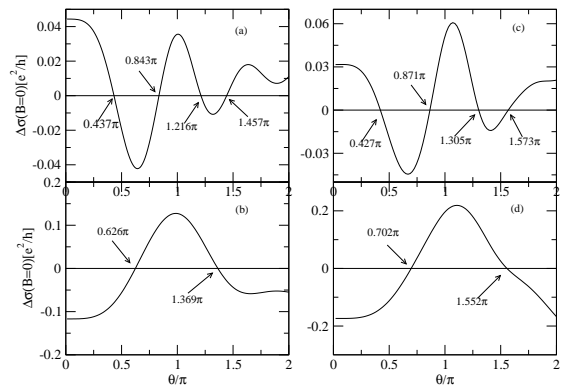}
\end{center}
\caption{ Numerical results for the amplitude of the AAS (a and c)
and the AB (b and d) oscillations at B=0 of quasi-1D square loop
system versus spin precession angle $\theta$. The system size is
N=12, M=6 for part (a), (b), and N=20, M=10 for part (c), (d). The
$\theta$ values of node positions are indicated in each graph. The
electron Fermi energy is E = -2.0.}
\end{figure}

\begin{figure}[ht]
\begin{center}
\includegraphics[height=8.5cm,angle=270]{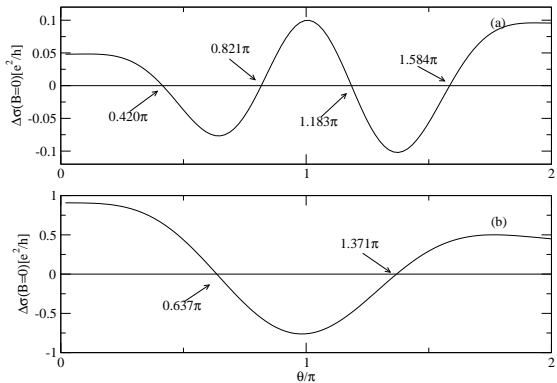}
\end{center}
\caption{ Numerical results for the amplitude of the AAS (a) and
the AB (b) oscillations at B=0 in an ideal 1D square loop system
versus spin precession angle $\theta$ with L=30. The $\theta$
values of node positions are indicated in the figure. The electron
Fermi energy is E =-1.7.}
\end{figure}

For the ideal 1D system, the numerical results exhibit that the
node positions $\theta^*$ of the amplitude of the AAS oscillation
are $0.420\pi, 0.821\pi, 1.183\pi, 1.584\pi$ etc (see Fig.4a), and
$\theta^*$ for the AB oscillation are $0.637\pi$ and $1.371\pi$,
etc (see Fig.4b). These values are calculated at electron Fermi
energy $E=-1.7$, which is close to the bottom of the band. Because
near the band bottom, the energy momentum dispersion relation is
almost quadratic, and the spin precession angle $\theta$ is
independent with $E$. So the node positions of $\theta$ remain
unchanged even at different Fermi levels $E$. These node positions
are in agreement with the previous theoretical results\cite{ref11}
or the Eq.(3) for the ballistic system, and the errors are within
$0.01\pi$ for all node positions $\theta^*$. This means that the
higher order tunneling processions (e.g. electron goes around the
loop multi-times) and the reflecting processions have limited
effect on the node positions $\theta^*$. Furthermore, our
numerical calculations also find that so long as the lengths of CW
and CCW paths are the same, the conductance is identical no matter
where the leads are connected to the perimeter of the loop
\cite{ref11}.

Next, let us discuss the numerical results of the quasi-1D system
(shown in Fig.3). The amplitudes $\Delta \sigma(B=0)$ of the AB
and AAS oscillations are oscillatory functions of the spin
precession angle $\theta$, which are similar with the ideal 1D
device. Moreover, it also clearly shows that $\Delta \sigma(B=0)$
is not a periodic function. This is contrast to an ideal 1D
device. In that case, $\Delta \sigma(B=0)$ versus $\theta$ is
approximately a periodic function with the period $\pi$ for the
AAS amplitude and $2\pi$ for the AB amplitude, like $A(\theta)$ in
previous theoretical calculation \cite{ref11} and Eq.(3) in this
paper. But this fact is consistent with the experimental results
\cite{ref8}.

Following, we focus on the node positions $\theta^*$ of the
oscillatory amplitudes, which have been given in Fig.3. These
values of $\theta^*$ have quite large difference with an ideal 1D
system, and the errors can be more than $0.1\pi$. For a loop with
wider arm, the discrepancy is larger. For example, for the arm's
width $M=10$, the second node position $\theta^*$ of the AB
oscillation is at $1.552\pi$ (see Fig.3d), which has an error of
$0.188\pi$ with the corresponding value $1.364\pi$ of the ideal 1D
device.

Therefore, we need to emphasize that if the width of the loop arm
in the device by Koga et al. has been taken into account, then the
node positions $\theta^*$ deviate from the ones for the ideal 1D
device. However, their experimental conclusion, that the spin
precession angle $\theta$ due to the Rashba SO interaction is
gate-controllable, remains valid. Furthermore the tunable range of
$\theta$ could be much larger than the value stated in their paper
since the node positions $\theta^*$ have larger deviations for
quasi-1D devices.

\begin{figure}[ht]
\begin{center}
\includegraphics[height=8.5cm,angle=270]{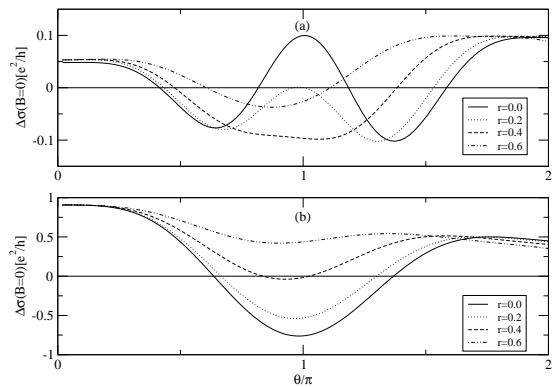}
\end{center}
\caption{ Numerical results for the amplitude of the AAS (a) and
the AB (b) oscillations at B=0 of an ideal 1D square loop system
versus Rashba dependent spin precession angle $\theta$ with L=30.
r is the ratio of Dresselhaus versus Rashba interaction strength
with $r \equiv \alpha_D/\alpha_R$. The Fermi energy is E=-1.7.}
\end{figure}

Up to now, we only consider the existence of the Rashba SO
interaction in the loop, and the Dresselhaus SO interaction has
been neglected. If there exists both Rashba and Dresselhaus SO
interactions, how are the node positions $\theta^*$ affected? In
fact, the 1D loop having two kinds of SO interactions has been
investigated by a very recent theoretical work by Ramaglia
\textit{et al.}\cite{ref18}. Their results show that Dresselhaus
SO interaction strongly shifts the nodes of Rashba-dependent
transmission in an ideal 1D model. Here we further investigate the
ideal 1D system with these two kinds of SO interaction. The tight
binding Hamiltonian of Dresselhaus SO interaction is given as:
\begin{equation}
H_{D}=\sum_{j}[i\hbar(\beta_{jy}a^{\dag}_{j} \sigma_y a_{j+\hat y}
-\beta_{jx} a^{\dag}_{j} \sigma_x a_{j+\hat x}) + h.c.]
\end{equation}
where $\beta_{j\tau}=\beta\cdot
e^{i\frac{e}{\hbar}\vec{A}\cdot\tau}$, $\beta=\alpha_D/2a$, and
$\alpha_D$ is the strength of Dresselhaus SO interaction. The
total Hamiltonian in the present system is the Hamiltonian in
Eq.(4) plus $H_D$. Then the amplitude of AAS and AB oscillation
can be solved by the same method as in the above.

In Fig. (5), we present the results of the amplitudes of AAS and
AB oscillations at B=0 for the exact 1D quare loop system as a
function of Rashba spin precession angle $\theta$ at different
values of Dresselhaus interaction. In this graph we could clearly
see that with the increasing of Dresselhaus SO interaction, the
node positions will not only vary, even the number of nodes may
change. For example in Fig. (5a), the number of nodes will
decrease from 4 to 3 at $r=0.2$ ($r\equiv \alpha_D/\alpha_R$), and
further to 2 at $r>0.2$. The same situation also happens in the AB
oscillation amplitude. These results are consistent with the
analytical results \cite{ref18}. Therefore, if both SO
interactions are presented in the system, using node positions to
determine the Rashba SO interaction strength is not very reliable.
However, from the oscillation of the amplitude AAS or AB versus
the gate voltage, it still clearly indicates that the strength of
SO interaction can be well tuned.

\begin{figure}[ht]
\begin{center}
\includegraphics[height=8.5cm,angle=270]{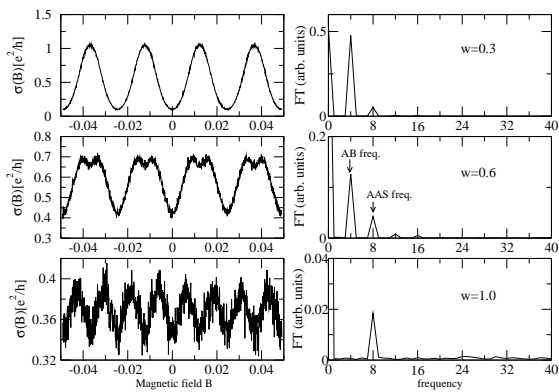}
\end{center}
\caption{Left panel: ensemble averaged conductance as a function
of magnetic field B at different disorder strength of the exact 1D
square loop system. Right panel: the corresponding Fourier
transform spectra of the conductance. The location of AB frequency
and AAS frequency is indicated in the graph. The parameters are
the Fermi energy E=-1.7, the system size L=16, and the Rashba
interaction $\lambda=0.1$.}
\end{figure}

Finally, we study how the node positions $\theta^*$ are affected
by the ensemble average. Notice that the device in the experiment
by Koga {\sl et al.} is a square loop array, not a single square
loop\cite{ref8}. So the ensemble average has to be investigated in
order to make a direct comparison with the experimental results.
The various ensemble averaging procedures have been studied in
investigating the persistent current in a closed mesoscopic ring
about ten years ago. Those include taking the ensemble average on
the number of particle, on the chemical potential, or on the
disorder \cite{addref5,addref6}. The present square loop array is
an open system, its chemical potential is determined by the leads,
but the disorder structure for each individual loop is random.
When giving the disorder structure and the chemical potential, the
number of electrons in the loop is fixed and its value can be
non-integer in an open system. So in the following we take the
average over the disorder ensemble\cite{addref6}. To consider the
existence of the disorder, the random lattice on site energy $w_j$
in Eq. (4) is generated by an uniformly distribution $[-w/2, w/2]$
with disorder strength $w$. All these conductance curves are
averaged over up to 2000 disorder realizations at $w\neq 0$. The
conductance as a function of magnetic field B at different
disorder strength $w$ are shown in Fig. (6) left panel, with their
corresponding FT (Fourier transform) spectra on the right panel.
With the increasing of disorder strength, both the conductance and
the conductance oscillation amplitude will decrease. This is due
to the fact that the electron will be localized at large disorder
strength. In particular, in the strong disorder case, AB
oscillation will disappear, but AAS oscillation still exists even
at the disorder strength $w=1.0$, because the AAS oscillation is
the interference between two time reversal pathes and it is
independent of the disorder structure. These results are in
agreement with the experimental results and the previous
theoretical predictions\cite{ref8,ref11}.

Next, we focus on the node positions $\theta^*$ after the disorder
ensemble average. Fig.7 shows the ensemble averaged AAS
oscillation amplitude at B=0 of an ideal 1D loop model at
different disorder strength $w$. $w=0$ is not shown in the graph,
because in that case the oscillation amplitude is much larger than
$w=0.6$ or $w=1.0$. It is clear that both the node positions of
$\theta$ and the oscillation shape is unchanged versus disorder.
These graphs demonstrate that with a little strong
spin-independent disorder and ensemble average, the AB oscillation
will disappear in the conductance. Meanwhile, the node positions
$\theta^*$ of AAS oscillation amplitude still remains the same.

\begin{figure}[ht]
\begin{center}
\includegraphics[height=8.5cm,angle=270]{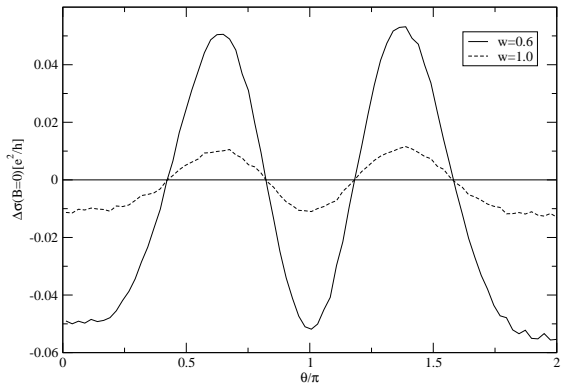}
\end{center}
\caption{The amplitude of the AAS oscillation at B=0 of an ideal
1D square loop system versus Rashba dependent spin precession
angle $\theta$ with L=20. The node positions are identical
compared with no disorder. The electron Fermi energy is E=-1.7.}
\end{figure}

In summary, by using the tight binding model and the
Landauer-B\"{u}ttiker formula, we numerically study the electron
transport through a quasi one dimensional square loop with the
Rashba spin-orbit interaction. The conductance $\sigma$ as a
function of the magnetic field $B$ is obtained, and exhibits the
Al'tshuler-Aronov-Spivak (AAS) and Aharonov-Bohm (AB)
oscillations. These oscillatory amplitudes are non-periodic with
the Rashba spin precession angle $\theta$. These results are in
agreement with the recent experiment by Koga \textit{et al.}
\cite{ref8}. To compare with an ideal 1D square loop device, the
node positions of the amplitudes of the AB and AAS oscillations
have some deviations. For a loop with a wider arm, the deviations
can be quite large. When under the influence of spin-independent
disorder and ensemble average, the AB oscillation will disappear
and only AAS oscillation survives in the conductance. In
particular, the node positions of the AAS oscillatory amplitude
are almost unchanged.

{\bf Acknowledgments:} We gratefully acknowledge financial support
from US-DOE under Grant No. DE-FG02-04ER46124 and NSF CCF-0524673.
QFS is supported by NSF-China under Grant Nos. 90303016, 10474125,
and 10525418. BC is supported by NSF-China under Grant Nos.
10574035 and 10274070.


\begin{references}
\bibitem{note11}
Here we consider that the Rashaba SO interaction only exists in
the loop and the leads are assumed to be without SO interaction.
In fact, if the leads have the SO interaction, our results, in
particular the node positions $\theta^*$, keep similar, because
the phase of the electron travelling through the loop is identical
regardless of the leads with or without the SO interaction.
\end{references}
\end{document}